\documentclass[twocolumn,aps,showpacs,fleqn,superscriptaddress]{revtex4}
\usepackage{epsfig}
\usepackage{graphicx}
\usepackage{amsfonts}
\usepackage{subfigure}
\usepackage[dvips]{color}
\usepackage{color}
\newcommand{\ig}{\includegraphics}
\newcommand{\ct}{\cite}
\newcommand{\bi}{\bibitem}
\newcommand{\de}{\delta}

\newcommand{\bra}{\langle}
\newcommand{\ket}{\rangle}

\newcommand{\be}{\begin{equation}}
\newcommand{\ee}{\end{equation}}
\newcommand{\ba}{\begin{eqnarray}}
\newcommand{\ea}{\end{eqnarray}}

\begin{document}

\title{Adiabatic multicritical quantum quenches: Continuously varying exponents depending on the direction of quenching}

\author{Victor Mukherjee}
\email{victor@iitk.ac.in}
\author{Amit Dutta}
\email{dutta@iitk.ac.in}
\affiliation{Department of Physics, Indian Institute of Technology Kanpur, Kanpur 208 016, India}

\begin{abstract}
We study  adiabatic quantum quenches across a quantum multicritical point (MCP) using a quenching scheme
that enables the system to hit the MCP along different paths. We show that the power-law scaling  of the defect
density with the rate of driving  depends non-trivially on the path, i.e., the exponent varies continuously with 
the parameter $\alpha$ that defines the path, up to a critical value $\alpha= \alpha_c$; on the other hand for $\alpha \geq \alpha_c$,  
the scaling exponent saturates to a constant value. We show that  dynamically generated  and {\it path($\alpha$)-dependent}
 effective critical exponents  associated with the quasicritical points lying close to the MCP (on the ferromagnetic side),
where the energy-gap is minimum, lead to this continuously varying exponent. 
The scaling relations are established using the integrable transverse XY spin chain and generalized to a MCP 
associated with a $d$-dimensional quantum many-body
systems (not reducible to two-level systems) using adiabatic perturbation theory.  
We  also calculate the effective {\it path-dependent}
dimensional shift $d_0(\alpha)$ (or the shift in center of the impulse region)  
that appears in the scaling relation for special paths lying entirely in the paramagnetic phase.  
Numerically obtained results are in good agreement with analytical predictions. 
\end{abstract}

\maketitle

Following the  Kibble-Zurek (KZ)\ct{kibble76,zurek96}  prediction of the scaling of the  density of defect in the final state of a
 quantum many-body system
following a slow  quench\ct{zurek05,polkovnikov05} across a quantum critical point (QCP)\ct{sachdev99,dutta96}, there has been 
an upsurge
in theoretical studies on quantum quenching across critical points 
\ct{damski05,polkovnikov07,mukherjee07,mukherjee08,sengupta08,santoro08,deng08,divakaran08}(for a review see \ct{dziarmaga09}).
 The KZ argument predicts that the scaling of 
the defect density ($n$) in the final state is universal and is given by $ n \sim  1/\tau^{\nu d/(\nu z+ 1)} $  
where $\tau$ is the inverse rate of driving across a QCP with the correlation 
length and dynamical exponents $\nu$ and $z$, respectively, and $d$ is the spatial dimension.
 The possibility of  the experimental verification of Kibble-Zurek scaling (KZS) in a
 spin-1 Bose condensate \ct{sadler06}, in ions trapped in optical lattices  \ct{duan03,bloch07}, 
and also in ultracold fermionic atoms in optical lattices \ct{lee09,dutta10} has paved the way for the 
above mentioned theoretical studies.

Although, the KZS for quenching through a quantum critical point is well-understood; the scaling of the defect
density following an adiabatic quantum quench across a quantum multicritical point (MCP) is relatively less studied. 
A non-KZS behavior ($n\sim 1/\tau^{1/6}$) of the density of defects (wrongly oriented spins) for quenching across the MCP of 
the spin-1/2 transverse XY chain was reported for the  first time in reference \ct{mukherjee07} which was later explained  in reference
\ct{divakaran09} 
introducing  an effective dynamical exponent $z_2(=3)$ for Jordan-Wigner solvable spin chains \ct{lieb61} reducible to a 
collection of decoupled two-level systems in the Fourier space and applying Landau-Zener (LZ) transition formula
 \ct{landau}.
  This argument was extended  to the  non-linear quenching of a general Hamiltonian in reference \ct{mondal09}.  
In a recent communication,  Deng $et~al$\ct{deng10}, attributed this anomalous scaling behavior to the
existence of quasicritical points close to a MCP where the energy gap is minimum and proposed a generic scaling for the multicritical quantum quenches
in terms of the effective critical exponents associated with these
quasicritical points. Studies on inhomogeneous
quantum transitions across a MCP has also been reported recently \cite{dziarmaga10}. But {\it how does the scaling exponent depend on 
the direction of approaching the MCP}?  In this work, we address this particular question proposing  a different quenching scheme
that enables the system to cross  the multicritical point along different directions and derive the corresponding KZS
which reduces to all the previous results in  appropriate limits.

At the outset, we summarize our main result using the example of a one-dimensional spin-1/2 transverse XY Hamiltonian 
\ct{barouch70} given by
\be
H = -\frac 1 {2} \sum_{j} \left({J_x \sigma^x_j \sigma^x_{j+1} + J_y \sigma^y_j \sigma^y_{j+1} + h\sigma^z_j}\right),
\ee
where $\sigma$'s are the Pauli spin matrices, $J_x$, $J_y$ are the interactions along $x$ and $y$, respectively, and $h$ denotes
the strength of the transverse field and henceforth, we set  $J_x +J_y$=1.  The parameter $\gamma=J_x -J_y$ 
is the measure of the anisotropy of the interactions and $J_x + J_y = \gamma = 1$ refers to  the transverse Ising limit \ct{pfeuty70}.
 The phase diagram of the Hamiltonian is well-known and shown in Fig.~1; we study
the quenching across the MCP (A) at $h=1,\gamma=0$ with the correlation length exponents and the dynamical exponents
 $\nu_1=1/2$ and $z_1=2$, respectively. 

We propose a generic quenching scheme where the parameters $h$ and
$\gamma$ are both varied simultaneously following the relation 
\be
h(t)= 1+ |\gamma(t)|^{\alpha}{\rm sgn(t)};\alpha >0 ~~{\rm and} ~~ \gamma(t)=-t/\tau
\ee

\noindent with time  $t$ going from $-\infty$ to $\infty$;
the system reaches the MCP A at $t=0$.  
The quenching path hits 
 the MCP vertically  (i.e., $|\partial \gamma/\partial h|_{\gamma \to 0} \to \infty)$ for $\alpha >1$ and horizontally
 for $\alpha <1$.   The parameter $\alpha$ introduces a non-linear temporal variation of $h$, thereby  determining
 the path of quenching across the MCP  as shown in Fig.~1; for $\alpha=1$ (path I), the MCP is approached linearly and hence the path is always equidistant 
from the Ising and Anisotropic critical lines. Similarly, if the parameters $h$ and $\gamma$  are  changed simultaneously  by
the same non-linear rate as
$\gamma(t)=h(t)-1=|t/\tau|^{\alpha}$ {\rm sgn(t)} as studied in  \ct{deng10},
the quenching path  once again happens to be equidistant from these two critical lines.  The same is not true for the quenching
scheme (2) if $\alpha \neq 1$; paths with $\alpha \to 0$ and $ \alpha \to \infty$ correspond to the passage across the anisotropic
and Ising critical lines, respectively.  For $\alpha=1$, the scheme (2) is the same as that  of reference \cite{deng10}.

 Our studies reveal  the existence of a critical
value $\alpha_c(=2$ for model (1))  for the quenching scheme (2). For $1 \leq \alpha <2$ , the defect density ($n_2$) 
scales as $n_2 \sim \tau^{-\alpha/6}$, i.e., the exponent varies continuously with the parameter $\alpha$ and saturates to the 
scaling $1/3$ for
$\alpha \geq 2$. We argue that the quenching path crosses the quasicritical points for $\alpha < 2$ but not 
for $\alpha >2$ when  we  recover the scaling relation valid for quenching the system  
across the MCP by linearly varying the
parameter $\gamma$  along the Ising critical line \cite{divakaran08} that amounts to setting
 $\alpha \to \infty$ in Eq.~(2).  We do also
derive similar scaling relation for the special path III (Fig.~1) and calculate the path-dependent dimensional
shift appearing in the scaling.  Finally, we extend our studies  
to a $d$-dimensional generic Hamiltonian 
using the adiabatic perturbation theory \cite{polkovnikov05} and establish the scaling relation,
$n_2\sim \tau^{-\alpha d \nu_2/[\alpha(z_2\nu_2+1)+z_1\nu_2(1-\alpha)]}$ where $z_1,\nu_1$ are the exponents
associated with the MCP while $z_2(\alpha)$ and $\nu_2(\alpha)=\nu_1 z_1/z_2(\alpha)$ are 
path dependent exponents associated with the quasicritical points. To the best of our knowledge this path-dependent
$z_2(\alpha)$ leading to a path-dependent continuously varying  exponent of KZS were never reported before. It needs
to be emphasized here that these path-dependent exponents are generally not useful in desrcribing  the equilibrium  quantum critical
behaviour though they do dictate the non-equilibrium dynamics of models as discussed above.
\begin{figure}
\ig[height=2.2in,width=3.2in,angle=0]{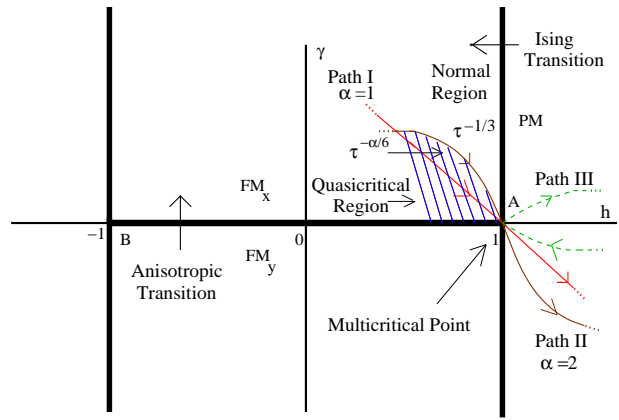}
\caption{(color online) The phase diagram of a one-dimensional $XY$ model in a transverse field. The vertical bold lines 
given by $h=\pm 1$ denote Ising transitions from the ferromagnetic phase to the paramagnetic phase. 
The horizontal bold line stands for anisotropic phase transition from a ferromagnetic phase with ordering in
the $x$-direction to a ferromagnetic phase with ordering in the $y$ direction. The multicritical points are shown by points 
$A$ ($h=1, \gamma=0$) and $B$($h=-1, \gamma=0$). We show different quenching paths corresponding to different values of $\alpha$;
path I (path II) is for $\alpha=1$ ($\alpha=2$). Path III corresponds to a  quenching scheme in which the system is always
 in the paramagnetic phase and touches the MCP at $t=0$. We show that in the shaded region there is a continuously
varying effective dynamical exponent and the exponent for defect density while for $\alpha \geq 2$, $n$ scales as $n \sim \tau^{-1/3}$.}
\end{figure}

\begin{figure}
\ig[height=1.9in,width=2.7in,angle=0]{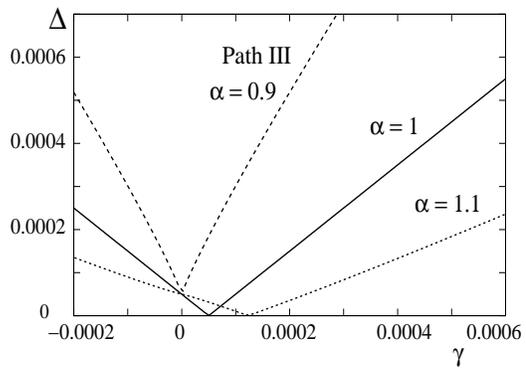}
\caption{The variation of energy gap as a function of $\gamma$ for $k=0.01$ close to MCP and also the quasicritical point 
for different $\alpha$.}
\end{figure}
It is well known that using the Jordan-Wigner transformation \ct{lieb61}, the Hamiltonian (1) can be reduced to direct product 
of decoupled $2\times 2$  Hamiltonians for each momentum $k$  given by  
\ba 
{H_k}=\left[ \begin{array}{cc} h+\cos k &  i\gamma \sin k\\
  -i\gamma \sin k& -(h+\cos k) \end{array}\right]\label{ham1}  
 \ea
and that the excitation gap is 
\ba
\Lambda_k = \sqrt{(h + \cos k)^2 + \gamma^2 \sin^2k}.
\ea
In the vicinity of the quantum MCP A,  with $\pi-k = k \to 0$,  Eq.~(3) can be simplified to 
\be
H_k = (h-1+k^2)\hat\sigma^z+ \gamma k \hat\sigma^x.
\ee

For the sake of comparison, let us briefly recall the  results of quenching  across the MCP A with $\gamma(t)=h(t)-1=|t/\tau|^{\alpha}{\rm sgn(t)}$
as studied in \ct{deng10} using the linearization process. 
The Hamiltonian (3) gets modified to
\ba
H_k &=& (|\frac t{\tau}|^{\alpha}{\rm sgn(t)}+ k^2)\hat\sigma^z+ |\frac t {\tau}|^{\alpha}{\rm sgn(t)} k \hat\sigma^x,\nonumber\\
&=& (|\frac t {\tau}|^{\alpha}{\rm sgn(t)}+ k^2)\hat\sigma^z+  k^3 \hat\sigma^x.
\ea
In deriving (6) an appropriate unitary transformation is used  so that
the time-dependence of the off-diagonal terms  is removed and the LZ  transition formula is directly applicable 
for $\alpha=1$ \ct{landau}.
We observe that the energy gap is minimum at a time $t_0$ so that $|t_0/\tau|^{\alpha}{\rm sgn(t)}+ k^2=0$;
this defines a quasicritical point at $t_0$ on the ferromagnetic side of the MCP (i.e., $t <0$). The minimum gap 
scales as $k^3$ so that we have an effective dynamical exponent $z_2=3$ even though the dynamical exponent at the MCP (i.e., $t =0$)
is still given by $z_1=2$; the equivalent correlation length exponent $\nu_2 = (\nu_1 z_1)/z_2=1/3$.  
We emphasize here that for this quenching scheme  \ct{deng10}, the exponent $z_2$ is fixed ($=3)$ irrespective
of the value of $\alpha$.

Since in the limit of $\tau \to \infty$, the maximum contribution to the defect comes from the vicinity of $t_0$, we linearize
the Hamiltonian  (6)  \ct{mondal09} around $t=t_0$ 
\ba
H_k = (\frac {t-t_0}{\tau_{\rm eff}})\hat\sigma^z+  k^3 \hat \sigma^x,
\ea
with $\tau_{\rm eff}= \tau k^{-2(\alpha -1)/\alpha}/\alpha$ and $\tau_{\rm eff}=\tau$ for $\alpha=1$. The higher order terms 
lead to a faster decay of the defect density and hence can be dropped in the limit of large $\tau$.
The excitation probability is now readily obtained using the
Landau-Zener formula $ p_k = \exp (-\pi k^{6}\tau_{\rm eff})$ which is integrated over all modes $k$
to derive the scaling of the defect density given as

\be
n = \int dk p_k\sim (\frac {\tau}{\alpha})^{-\alpha/[6\alpha+2(1-\alpha)]}\sim (\frac {\tau}{\alpha})^{-\alpha/[2(2 \alpha+1)]}.
\ee
It is straight forward to generalize to a  $d$-dimensional quantum MCP  with $\nu_1 z_1=1$ described the
Hamiltonian
\ba
H_k &=& (\lambda_1 +k^{z_1})\hat \sigma^z+ \lambda_2 k^{\beta} \hat\sigma^x,\\
&=&(|\frac t {\tau}|^{\alpha}{\rm sgn(t)}+ k^{z_1})\hat\sigma^z+ |t/\tau|^{\alpha}{\rm sgn(t)} k^{\beta} \hat \sigma^x.
\ea
where the MCP is reached at time $t=0$ due to the simultaneous variation of two parameters $\lambda_1$ and $\lambda_2$
under the quenching scheme $\lambda_1 = \lambda_2 =|t/\tau|^{\alpha} {\rm sgn (t)}$.  Comparison with the Hamiltonian (5) gives
  $\lambda_1=h-1$, $\lambda_2 = \gamma$, $z_1=2$ and $\beta=1$. Linearizing around $t_0$ and using 
$\tau_{\rm eff}= \tau k^{-z_1(\alpha -1)/\alpha}/\alpha$, we similarly get
\be
n \sim \tau^{-\alpha d/[2z_2\alpha+z_1(1-\alpha)]}; z_2 =z_1+\beta
\ee

\begin{figure}
\ig[height=1.9in,width=2.8in,angle=0]{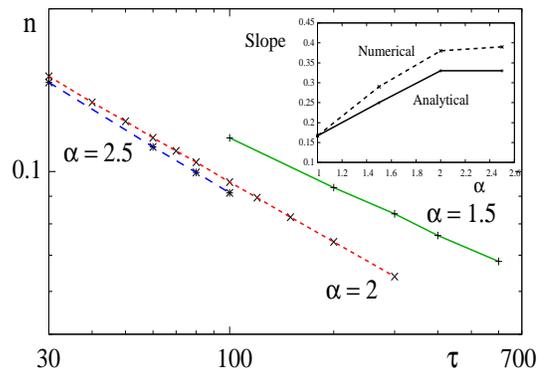}
\caption{(color online) The defect density obtained by numerical integration of the Schr\"odinger equation for path $\alpha=1.5$ (slope $\approx 0.29$,
 green line), $\alpha=2$ (slope $\approx 0.38$, red line) and $\alpha=2.5$ (slope $\approx 0.39$, blue line). In the inset, we show a comparison between
numerically obtained (dashed line) and analytical obtained (solid line) slopes for different $\alpha$. The numerical results are 
in good agreement with the analytical predictions; the observed  small mismatch is
 due to neglecting the higher order corrections in the process of linearizing.}
\end{figure}

\begin{figure}
\ig[height=1.9in,width=3.0in,angle=0]{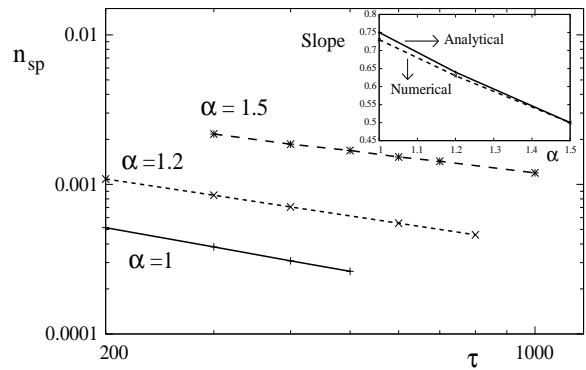}
\caption{Defect density obtained by numerical integration of the Schr\"odinger equation for special
 path III; $\alpha=1$ (slope $\approx 0.73$), $\alpha=1.2$ (slope $\approx 0.63$) and $\alpha=1.5$ (slope $\approx 0.5$). Inset shows the comparison
 between numerically obtained and analytical obtained slopes for different $\alpha$. The numerical results are in good agreement with the analytical
 predictions.}
\end{figure}

Although scaling given in Eq. ~(11) matches identically with the scaling relation Eq.~(4) of reference \cite{deng10} for $\alpha=1$
(with $d_0=0$ and $\nu_1z_1 = \nu_2 z_2 =1$),
there is a  small difference  for $\alpha \neq 1$. This mismatch
stems from the fact that we have used linearization of the Schr\"odinger equations in the vicinity
 of $t_0$ (or $\tau_{\rm eff}$) which necessarily yields a term $z_1(1-\alpha)$ in the KZS  for $\alpha \neq 1$. 
We do also note that for $\alpha <1$,  $\tau_{\rm eff} \to 0$ in the limit $k \to 0$ rendering
the process of linearization inappropriate. In the following, we shall restrict our analytical calculations 
to $\alpha \geq 1$.

Now for our quenching scheme (2), the Hamiltonian (3) takes the form 

\be
H_k = (|\frac {t}{\tau}|^{\alpha}{\rm sgn(t)}+ k^2)\hat\sigma^z+ \frac t {\tau}  k \hat\sigma^x.
\ee
We note that the energy gap of Hamiltonian (12)
is minimum at a quasicritical point for $t=t_0$ where $|t_0/\tau|^{\alpha}{\rm sgn(t)}+ k^2=0$ only if $\alpha <2$ while
for $\alpha >2$, the minimum gap occurs right at the MCP (i.e., at $t=0$) (see figures (1) and (2)). Linearizing 
around $t=t_0$, we get 
\ba
H_k  =\frac {t-t_0}{\tau_{\rm eff}}\hat\sigma^z+  k^{2/\alpha +1} \hat\sigma^x
\ea
with $\tau_{\rm eff}= \tau k^{-2(\alpha -1)/\alpha}/\alpha$. Here, we have ignored the time dependence of the off-diagonal term
in favor of the minimum value at $t=t_0$  scaling as $k^{2/\alpha +1}$ which is appropriate for $1 \leq \alpha <2$ but not for $ \alpha \geq 2$
since $ \tau_{\rm eff}/(\tau/k) \sim k^{(2-\alpha)/\alpha} \to 0$ as $k\to 0$ if $\alpha <2$. This implies that $\tau_{\rm eff}$ 
dictates the scaling behavior of the defect for the dynamics of Eq.~(12) for $1 \leq \alpha < 2$ while for $\alpha \geq 2$ a crossover
to the scaling behavior dictated by $k t/\tau$ is expected.  Further, the exponent
 $z_2 = z_2 (\alpha) = 2/\alpha +1$ depends on $\alpha$ though $z_1$ is completely independent of the path and fixed  ($ =2$).
We emphasize that the existence of the path dependent $z_2$ is the crux of the quenching scheme (2) which leads to
striking consequences shown below. We already note that $\alpha=2$ is a marginal case where $z_2$ becomes equal to $z_1$

Application of the Landau-Zener formula to Eq.~(13) similarly yields a scaling relation for the defect density $(n_2)$  
for the quenching
scheme (2) given by
\be 
n_2 = \int dk \exp (-\pi  k^{\frac {4}{\alpha}+2} \tau_{\rm eff})  \sim (\frac{\tau}{\alpha})^{-\alpha/6}
\ee
so that the exponent varies continuously  with the parameter $\alpha$. In the limit $\alpha=1$, we recover the well known
scaling $n \sim \tau^{-1/6}$ \ct{divakaran09} while for $\alpha \geq 2$, we get   $n \sim \tau^{-1/3}$ \ct{divakaran08,deng08}
the system  hits the MCP vertically and 
does no longer  cross any quasicritical point  along the path and hence  the quenching scheme  boils down to driving the  system across the MCP along the Ising critical line \ct{divakaran08} when  the term 
$|t/\tau|^{\alpha}$ in Eq.~(12) can essentially be ignored. The defect density as a function $\tau$ for 
different $\alpha$ obtained by numerical
 diagonalization of the  Schr\"odinger equations in case of full path is shown in  Fig.~3, and the results are in good agreement with 
the prediction of Eq.~(14). We note that the linearization process retains only the
 slowest $\tau$-dependence of $n$, while contributions
from the higher order terms decay must faster. However, this can be numerically observed only in the limit of sufficiently large $\tau$. We reiterate that
 the scaling relation given in (14) is inappropriate for $ \alpha <1$
though our numerical results do indicate the existence of continuously varying exponent that decreases with decreasing $\alpha$.

Generalizing to the  $d$-dimensional Hamiltonian (9) with $\tau_{\rm eff}= \tau k^{-z_1(\alpha -1)/\alpha}/\alpha$,
 we get the relation 
\be 
n_2 \sim \tau^{-\alpha d/[2z_2 \alpha +z_1 (1-\alpha)]}
\ee
Although 
relation (15) closely resembles (11), it is to be noted
that in our quenching scheme
the exponent  $z_2 (=z_1/\alpha + \beta)$
and hence, the scaling exponent  depends on $\alpha$ up to $\alpha=\alpha_c$ when $z_1(\alpha_c)=z_2(\alpha_c)$ 
i.e., $\alpha_c= z_1/(z_1 -\beta)$.

Finally, we consider the quenching scheme 
\be
h(t)= 1+|\gamma|^{\alpha} ~~{\rm and}~~ \gamma=t/\tau,
\ee
 in which the system  lies in the paramagnetic phase throughout 
the quenching process and touches the MCP at $t=0$ (see path III in Fig.1). The case $\alpha=1$ was studied in reference
\ct{deng10} and a scaling relation of the form $n \sim \tau^{-3/4}$ was obtained. This anomalous  scaling behavior was justified using 
the argument  of  a dynamical shift  of the center of the impulse region \ct{deng10} which makes it asymmetric as opposed to the
 conventional KZS such that non-critical energy states contribute to the defect
 with $p_k \sim k^{d_0(\alpha)}$ where $d_0(\alpha)$ is the additional effective dimension which eventually appears in
the KZS. In a similar spirit, we use
the the  LZ formula \ct{vitanov99} for  Hamiltonian (13) for the half path  so that the singularity at $t=0$ is avoided; 
this provides the correct scaling for the full path since the symmetry of the initial and final states are the same. 
Expanding  the transition probability (Eq.~(7) of \ct{vitanov99}) in Taylor series around $T_f=0$, we get 
dominant contribution given as $ p_k=|\omega/T_f|^2$ where the  dimensionless
time $T_f = (k^{2/\alpha} \tau)/\sqrt{\tau_{\rm eff}}$ and the dimensionless coupling 
$\omega = k^{(2/\alpha +1)}\sqrt{\tau_{\rm eff}}$.
Using $\tau_{\rm eff}=\tau k^{-2(\alpha -1)/\alpha}/\alpha$, we get $p_k= (1/\alpha^2) k^{2(2-\alpha)/\alpha}$.
The approximation is valid up to $k=k_{\rm max}$ for which $T_f=1$ given by $k_{\rm max} \sim \tau^{-\alpha/2(\alpha+1)}$. 
The scaling of the defect density ($n_{\rm sp}$) for these special paths is hence given by
\be
n_{\rm sp} =\int^{k_{\rm max}} dk p_k \sim k_{\rm max}^{\frac{4-\alpha}{\alpha}} \sim \tau^{-\frac {4- \alpha}{2(1+\alpha)}},
\ee
where in the limit $\alpha=1$, we recover the scaling $\tau^{-3/4}$
as in \cite {deng10}. 
We conclude that for quenching scheme (16) the dimensional shift  $d_0(\alpha)= 2(2-\alpha)/\alpha$, varies continuously with $\alpha$;
 $d_0(\alpha)=2$  for $\alpha=1$ and vanishes
for $\alpha \geq \alpha_c (=2)$ when once again the impulse region becomes symmetrical. Interestingly, for these special 
paths also we obtain a  continuously varying  exponent which saturates to $1/3$ in the limit $\alpha \to \alpha_c$.  Fig.~4 shows the
 defect density as a function of $\tau$ for different $\alpha$ for the special path III, as obtained from numerical integration.

We shall now derive  the  generalized scaling relation for   a quantum system with $\nu_1 z_1 \neq1$
using adiabatic perturbation theory \ct{polkovnikov05}. We note from  Eq.~(13) that the dynamics across a MCP
 can be described as a linear quench with a different
$\tau_{\rm eff}$ for each $k$ mode across a quasicritical point which provides the most dominant
contribution to the defect. 

For a general Hamiltonian parametrized by $\lambda = t/\tau$, we can write the wave function as $\psi = \sum_p{a_p (t) \phi(\lambda)}$. In the above 
basis the Schrodinger equation becomes
\ba
\frac{i}{\tau} \frac{d a_p(t)}{d\lambda} + \frac{i}{\tau}\sum_{q}{a_{q}(t) \langle p| \frac{d}{d\lambda}|q\rangle} = E_p (\lambda) a_p (t),
\ea
where $E_p (\lambda)$ is the eigenenergy of $H(\lambda)$. A unitary transformation
\ba
a_p(t) = \tilde{a}_p (\lambda) e^{-i \tau \int^{\lambda} E_p (\lambda^{'}) d\lambda^{'}}
\ea
transforms eq.~(18) to
\ba
\frac{d \tilde{a}_p (\lambda)}{d \lambda} = - \sum_{q} {\tilde{a}_q (\lambda)\langle p| \frac{d}{d\lambda}|q\rangle e^{i \tau \int^{\lambda} (E_p (\lambda^{'}) - E_q (\lambda^{'}))  d\lambda^{'}}}.
\ea
Now taking into account that initially the system was in the ground state, which makes a single term dominate the sum in eq. (20),
 and expressing the above equations in the momentum basis, we get for our case,
\be n \simeq \int \frac{d^d k}{(2 \pi)^d} \Big| \int_{-\infty}^\infty d
\lambda \bra \vec{k}|\frac{\partial}{\partial \lambda}|0\rangle  e^{i \tau_{\rm eff}
(|\vec{k}|;\alpha) \int^\lambda d \lambda'\de E_{\vec {k}} (\lambda')} \Big|^2,
 \ee
 where near a quasicritical point, $\de E_{\vec {k}} (\lambda) \simeq |\lambda|^{
z_2 \nu_2} F'(|\vec {k}|^{z_2}/|\lambda|^{z_2 \nu_2})$. Here $\lambda$ denotes the deviation
from the quasicritical point and $F'(x) \sim x$ for
large $x$. Again, we have  $\bra \vec {k}|\frac{\partial}{\partial \lambda} |0\ket =
\frac{1}{\lambda}G(k /\lambda^{\nu_2})$  where
$G(0)$ is a constant.
 Using these relations, defining 
$\lambda'=\lambda/|\vec k|^{1/\nu_2}$ and eventually rescaling
$|\vec{k}| \to |\vec{k}|
\tau^{\alpha \nu_2/\left[\alpha (z_2\nu_2+1) + z_1 \nu_2 (1 -
\alpha)\right]}$, we  get
\be n_2 ~\sim ~\tau^{-d \alpha \nu_2
/\left[  \alpha( z_2 \nu_2+1)+z_1 \nu_2 (1-\alpha) \right]},
 \ee 
which reduces to Eq. (15) for $ \nu_2 z_2
=1$. Of course, the scaling relation (22) holds up to a value
of $\alpha=\alpha_c$ such that $z_1(\alpha_c)=z_2(\alpha_c)$ and for $\alpha \geq \alpha_c$ 
gets modified to $ n ~\sim ~\tau^{-d \alpha_c \nu_1/
\left[\alpha_c+ \nu_1 z_1 \right]}$. For special paths (e.g., path III), we note that the  minimum gap 
is given by exponents associated with the MCP  and derive
the generalized scaling relation  given by
\be n_{\rm sp} \sim \tau^{-(d + d_0 (\alpha))\alpha \nu_1/\left[ \alpha z_1 \nu_1+1 \right]},
\ee 
where we have  included the dimensional shift $d_0 (\alpha)$
which varies continuously with $\alpha$ for $\alpha < \alpha_c$, while for $\alpha > \alpha_c$ the scaling saturates
to $n \sim \tau^{-\alpha_c \nu_1 d/(\alpha_c \nu_1 z_1 + 1)}$.

The main results of the paper are summarized in Eqs.~(14,15,17) and (22-23).
In conclusion, we propose a different  quenching scheme across the MCP and derive a KZS with an exponent continuously varying
with the path as a consequence of path dependent effective dynamical exponents associated with the quasicritical points.
 This continuously varying exponent  up to 
a critical value of $\alpha$ is  the cardinal feature of this paper. Also, our studies establish that unlike the conventional
KZS, the path-dependent effective exponents which usually do not  appear in studies of the equlibrium quantum critical behavior,
do appear in the scaling relation of the defect density in special situations like quenching across a multicritical
point.



We acknowledge Gerardo Ortiz, Anatoli Polkovnikov, Giuseppe Santoro
and Diptiman Sen for helpful comments and discussions. We also acknowledge the hospitality of Abdus Salam ICTP and SISSA, Trieste,
 where the present version of the paper was written. AD acknowledges CSIR, New Delhi for financial support through
the project SPO/CSIR/PHY/2010072.


\end{document}